\documentclass{vgtc}                          

\ifpdf
  \pdfoutput=1\relax                   
  \usepackage{graphicx}                
  \DeclareGraphicsExtensions{.pdf,.png,.jpg,.jpeg} 
\else
  \usepackage{graphicx}                
  \DeclareGraphicsExtensions{.eps}     
\fi%

\graphicspath{{figures/}{pictures/}{images/}{./}} 

\usepackage{microtype}                 
\PassOptionsToPackage{warn}{textcomp}  
\usepackage{textcomp}                  
\usepackage{mathptmx}                  
\usepackage{times}                     
\usepackage{cite}                      
\usepackage{tabu}                      
\usepackage{booktabs}                  
\usepackage{todonotes}
\usepackage{soul}
\usepackage{balance}

\usepackage{xcolor,colortbl} 

\sethlcolor{cyan}
\renewcommand\hl[1]{#1}

\onlineid{1284}

\vgtccategory{Research}

\vgtcinsertpkg

\title{Breaking the Screen: Interaction Across Touchscreen Boundaries in Virtual Reality for Mobile Knowledge Workers}

\author{
 Verena Biener$^{1}$
\and Daniel Schneider$^{1}$ 
\and Travis Gesslein$^{1}$
\and Alexander Otte$^{1}$
\and Bastian Kuth$^{1}$
\and Per Ola Kristensson$^{3}$
\and Eyal Ofek$^{2}$
\and Michel Pahud$^{2}$
\and Jens Grubert$^{1}$\thanks{contact author: jens.grubert@hs-coburg.de}
}
\vspace{-0.15cm}
\affiliation{\scriptsize $^{1}$Coburg University of Applied Sciences and Arts  $^{2}$Microsoft Research  \\ $^{3}$University of Cambridge}

\teaser{
\vspace{-0.3cm}
\centering 
  \includegraphics[width=0.9\columnwidth]{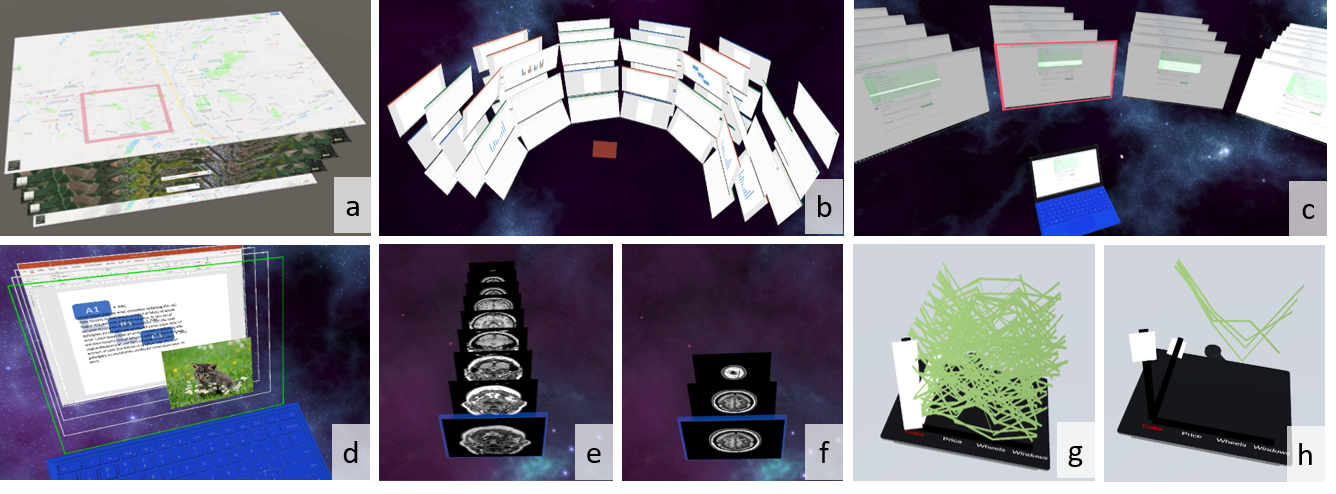}
  \caption{
  Mobile knowledge worker applications using a touchscreen in VR implemented in this paper. a: map navigation, b: window manager, c: code version control, d: presentation editor, e-f: medical imaging, g-h: information visualization.}
  \label{fig:teaser}
}

\abstract{

Virtual Reality (VR) has the potential to transform knowledge work. One advantage of VR knowledge work is that it allows extending 2D displays into the third dimension, enabling new operations, such as selecting overlapping objects or displaying  additional layers of information. On the other hand, mobile knowledge workers often work on established mobile devices, such as tablets, limiting interaction with those devices to a small input space. This challenge of a constrained input space is intensified in situations when VR knowledge work is situated in cramped environments, such as airplanes and touchdown spaces.

In this paper, we investigate the feasibility of interacting jointly between an immersive VR head-mounted display and a tablet within the context of knowledge work. Specifically, we 1) design, implement and study how to interact with information that reaches beyond a single physical touchscreen in VR; 2) design and evaluate a set of interaction concepts; and 3) build example applications and gather user feedback on those applications.

} 

\begin{document}


\firstsection{Introduction}

\maketitle


Recent progress in virtual reality (VR) technology makes it possible to provide knowledge workers with a portable virtual office. This office can provide many potential advantages for mobile work, such as 1) providing a well-illuminated, private environment with wide display areas regardless of the physical surroundings; 2) enabling a virtual recreation of a consistent spatial workspace; and 3) relieving the user of physical limitations, using large and even distant displays while the user is remaining seated and the hands are resting on a table \cite{grubert2018office}.

The vision of a spatial user interface supporting knowledge work has been investigated for many years (e.g.~\cite{raskar1998office, rekimoto1999augmented}). However, the recent emergence of consumer VR headsets now make it feasible to explore the design of deployable user interface solutions. 

However, while VR is promising for mobile office work it also brings its own additional challenges. For example, mobile workers often work in front of touchscreens, such as tablets or laptops. Within this work, we propose to extend the limited output and input space of such configurations by utilizing VR head-mounted displays (HMDs). Specifically, recent  HMDs with inside out camera-based sensing allow for pose tracking of the user's head and hands. Through access to HMDs cameras, spatial tracking of further objects, such as screens or keyboards become feasible.


Given these opportunities, this paper explores how to leverage the large display area and additional three-dimensional display volume around the user afforded by an HMD while using established capacitive touch interfaces provided with tablets and some laptops.

This paper addresses this scenario with the following contributions. First, we explore and evaluate how to spatially arrange and manipulate information within the joint interaction space of HMD-tablet interaction. Second, we design, implement and evaluate six applications (window manager, code version control, parallel coordinates exploration, map navigation, volumetric image viewer, and a virtual PowerPoint), see Figure \ref{fig:teaser}.

\section{Related Work}


The work in this paper is underpinned by the following research areas: 1) mixed reality (MR) for knowledge work; 2) information windows in spatial environments; and 3) spatial interaction.

\subsection{Mixed Reality for Knowledge Work}

Recent prior work has begun exploring how to support knowledge work using MR  \cite{grubert2018office, ruvimova2020transport, guo2019mixed}. Early work investigated the use of projection systems to extend physical office environments for interaction with physical documents  (e.g., \cite{wellner1994interacting, kobayashi1998enhanceddesk, rekimoto1999augmented, pinhanez2001everywhere}). VR and AR HMDs have also been explored for interacting with physical documents (e.g., \cite{grasset2007mixed, li2019holodoc}). Most prior work has focused on annotating documents displayed on a 2D surface while this work investigates the use of space surrounding a planar piece of information.


In addition to enhanced document interaction, prior work has also explored remote presence applications (e.g., \cite{raskar1998office, pejsa2016room2room}). There is a number of publications investigating the use of VR in desktop-based environments for tasks such as text entry (e.g., \cite{mcgill2015dose, knierim2018physical, grubert2018text}), system control \cite{zielasko2019passive, zielasko2019menus} and visual analytics \cite{wagner2018virtualdesk}. B{\"u}schel et al.~\cite{buschel2018interaction} surveyed a wide range of immersive interaction techniques for visual analytics. Previous research on productivity desktop-based VR has concentrated on the use of physical keyboards \cite{schneider2019reconviguration}, controllers and hands \cite{kry2008handnavigator, zielasko2019passive}, and, recently, tablets \cite{surale2019tabletinvr}. The closest work to this paper is by Surale et al.~\cite{surale2019tabletinvr}, which focuses on spatial manipulation and creation of 3D objects. In contrast, we use the tablet for information management on virtual displays.

Complementary to this prior research, this paper aims to support mobile knowledge workers by extending commonly used tools such as tablets and notebooks through HMDs. 

\subsection{Information Windows in Spatial Environments}
In 1993, Feiner et al. \cite{feiner1993windows} introduced head-surrounding and world reference frames for positioning 3D windows in VR. In 1999, Mark Billinghurst and Thad Starner \cite{billinghurst1999wearable} introduced the spatial display metaphor, in which information windows are arranged on a virtual cylinder around the user. 

Since then, virtual information displays have been explored in various reference systems, such as world-, object-, head-, body- or device-referenced~\cite{laviola20173d}.  Specifically, interacting with windows in body-centered reference systems~\cite{wagner2013body} has attracted attention, for instance to allow fast access to virtual items \cite{li2009virtual, chen2012extending}, mobile multi-tasking \cite{ens2014personal, ens2014ethereal} and visual analytics~\cite{ens2016spatial}. Gugenheimer et al.~\cite{gugenheimer2016facetouch} introduced face touch, which allows interacting with display-fixed user interfaces (using direct touch) and world-fixed content (using raycasting). \hl{Yu et al.}~\cite{yu2019depthmove} \hl{investigated the use of body motions for switching interfaces in VR.} Lee et al.~\cite{lee2018projective} investigated positioning a window in 3D space using a continuous hand gesture. Petford et al.~\cite{petford2018pointing} compared the selection performance of mouse and raycast pointing in full coverage displays (not in VR). Recently, Jetter et al.~\cite{jetter2020vr} proposed to interactively design a space with various display form factors in VR. \hl{Wagner et al.~proposed a desk metaphor for controlling visual analytics that reappropriates a physical desk in VR} \cite{wagner2018virtualdesk}.

Prior work has also explored how to support user interaction across HMDs and mobile and wearable touch displays.  Grubert et al.~\cite{grubert2015multifi} presented body-aligned, device-aligned, and side-by-side modes for interacting between a touch display (smartphone or smartwatch) and an optical see-through HMD. Similar explorations have followed suit using video-see-through HMDs \cite{normand2018enlarging}, an extended set of interaction techniques \cite{zhu2020bishare}, using smartwatches \cite{wenig2017watchthru, luwatchar, wolf2018performance}, or with a focus on understanding smartphone-driven window management techniques for HMDs \cite{ren2020understanding}. In a similar vein, prior work has studied the interaction across HMDs and stationary displays, such as  tabletops \cite{serrano2015gluey, butscher2018clusters, reipschlager2019designar}.

Most prior work relate to the issue of field of view, that is, how to display and access more information by spreading it around the user in multiple windows. In this research, we are additionally interested in extending the information display in the depth direction. Such a display is suited for displaying different views of information (or layers) that are semantically connected by their 2D location. On the other hand, we use a very limited input space: While most referred prior work span all the angular range of the display as an input space, we  only use the interaction space on or near the tablet in order to support interaction in constrained physical spaces \cite{grubert2018office, mcgill2019challenges}. 

\subsection{Spatial Interaction}

A large number of techniques for selection, spatial manipulation, navigation, and system control have been proposed to support spatial user interfaces~\cite{laviola20173d}. Regarding object selection, Argelaguet et al.~\cite{argelaguet2013survey} presented a survey on 3D selection techniques. For a recent survey on 3D virtual object manipulation, we refer to Mendes et al.~\cite{mendes2019survey}. Finally, recent surveys \cite{brudy2019cross, grubert2016challenges} have extensively reviewed spatial interaction across mobile devices, mostly in non-VR settings.

In addition to unimodal techniques the combination of touch with mid-air has attracted attention from researchers. For example, outside of VR, M{\"u}ller et al.~\cite{muller2014mirrortouch} investigated the use of touch and mid-air interaction on public displays, Hilliges et al.~\cite{hilliges2009interactions}  studied tabletop settings. Several researchers have proposed to use handheld touchscreens in spatial user interfaces for tasks such as sketching, ideation and modeling (e.g., \cite{dorta2016hyve, ramanujan2016mobisweep, huo2017window, arora2018symbiosissketch, gasques2019pintar, drey2020sketching}), navigation of volumetric data \cite{song2011wysiwyf}, 3D data exploration \cite{lopez2015towards} and system control \cite{bowman2001design}. Spatial manipulation has mostly been studied in single-user settings (e.g., \cite{szalavari1997personal, mossel20133dtouch, marzo2014combining, babic2018pocket6, liang2013investigation, katzakis2015inspect, surale2019tabletinvr}) but also in collaborative settings \cite{grandi2017design}. 

Evolving from the magic lens \cite{bier1993toolglass} and tangible interaction concepts \cite{ullmer1997metadesk}, tangible magic lenses allow users to access and manipulate otherwise hidden data in interactive spatial environments. A wide variety of interaction concepts have been proposed within the scope of information visualization (e.g., recent surveys \cite{tominski2014survey, tominski2017interactive}). Both rigid shapes (e.g., rectangular \cite{spindler2009paperlens}) or circular \cite{spindler2010tangible} and flexible shapes (e.g., \cite{steimle2013flexpad}) have been used, as well as various display media (e.g., projection on cardboard \cite{spindler2009paperlens, chan2012magicpad}), transparent props \cite{schmalstieg1999using, brown2006magic}, handheld touchscreens \cite{grubert2015utility, leigh2015thaw}, or virtual lenses \cite{oh2006user, looser20073d}. In addition, the combination of eye-gaze with other modalities such as touch \cite{pfeuffer2015gaze, pfeuffer2016gaze}, mid-air gestures \cite{pfeuffer2017gaze+, schweigert2019eyepointing, ryu2019gg} and head-movements  \cite{ kyto2018pinpointing, sidenmark2019eye,  sidenmark2020bimodalgaze} has been recently investigated for interaction in spatial user interfaces. For a recent survey on gaze-based interaction in AR and VR, see Hirzle et al.~\cite{hirzle2019design}.

We draw on these rich sources of interaction ideas and adopt techniques in the context of VR interaction with touchscreens for mobile knowledge workers. Our work complements multimodal techniques combining touch and mid-air \cite{hilliges2009interactions, muller2014mirrortouch}, gaze-based techniques \cite{pfeuffer2015gaze} and  ideas for combining HMDs with touchscreens \cite{grubert2015multifi, normand2018enlarging, zhu2020bishare} through novel techniques for accessing virtual windows around or behind a physical touchscreen.

\section{Design Space}
\label{sec:design}


Currently most knowledge workers' applications are designed for 2D displays, which are the vast majority of displays in the world. VR and AR displays enable mobile workers to take their displays with them on the go. In addition, since HMDs are stereoscopic, they enable information presentation beyond the 2D display and the ability to manipulate this information by spatial manipulation. Researchers have already proposed various schemes for arranging (planar) information in a spatial context relative to the user. Specifically, prior work has proposed various reference systems, such as world-, object-, head-, body- or device-referenced \cite{bowman20043d}, spatial window layouts in these reference frames, for example, scaffolding geometry such as spheres or cylinders \cite{ens2014ethereal, ens2016spatial}, and input modalities to access and manipulate information in these spaces (such as touch, gaze, mid-air interaction or multimodal interaction, c.f. \cite{wagner2013body, hirzle2019design, buschel2018interaction}).

The guiding principle in this paper is that an information touchscreen display can break out of the screen in VR and transition from 2D to 3D---yet still be controlled by manipulating the original touchscreen. This is in part motivated by prior work reviewed in the previous section and in part motivated by the fact that a touchscreen allows the user to provide precise 2D input while displaying information in an HMD provides the ability to display information in 3D. This screen-centric manipulation using touch can be complemented with additional modalities, such as gaze tracking.


Two important aspects arising in this context are how to \textit{spatially arrange} information relative to the touchscreen and how to \textit{map the input from the physical touchscreen to the information in the virtual world}, given the joint input capabilities of 2D touch sensing (on the screen) and 3D spatial manipulation (around the screen). 

\subsection{Spatial Arrangement}

\begin{figure*}[t!]
	\centering 
	\includegraphics[width=1.6\columnwidth]{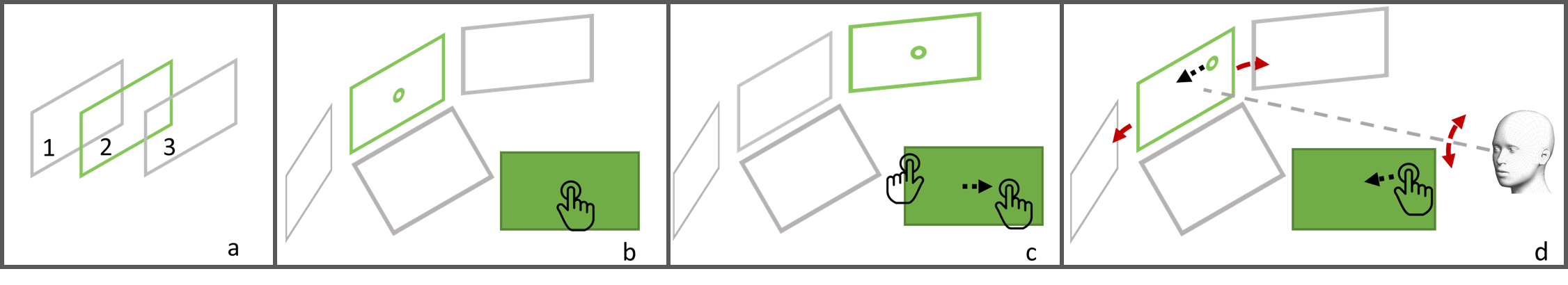}
	\caption{a: Virtual displays (grey) arranged relative to the physical screen (green): behind (1), with (2), in front of (3). b and c: Changing between pointing on a single screen and switching between screens using a bimanual technique. b: If only a single finger is used on the touchscreen (filled green rectangle), it controls the cursor on the active screen (green boundary) with a suitable control-display gain. c: The screens are switched by pressing with a finger of the non-dominant hand on the bezel of the touchscreen and moving the finger of the dominant screen towards the second screen (in this case the right screen) using a second CD-ratio. d: Changing between pointing on a single screen and switching between screens using a combination of touch and gaze. The finger on the touch screen (filled green rectangle) controls the cursor on the active screen (green boundary). Gaze provides the region of interest. If the user gazes to the side, up or down, the corresponding screen gets activated (with appropriate temporal and spatial thresholds). }
	\label{fig:id}
\end{figure*}

Prior work has investigated options on how to align multiple 2D information windows around the user (e.g.,  \cite{ens2014ethereal, ens2014personal}). One common approach is to arrange windows on a scaffolding geometry, such as spheres or cylinders \cite{billinghurst1999wearable}
 (see Figure \ref{fig:id}, b-d). In a spatially unconstrained environment, accessing this information is often realized using virtual hand or raycasting techniques.
While using such extended display areas is possible, we should be aware that in constrained spaces, such as on an airplane or in touchdown spaces, \hl{it may not be usable for interaction} as ample spatial movements can be unsuitable either due to social acceptability \cite{ahlstrom2014you, williamson2019planevr} or simply due to lack of physical space to operate \cite{grubert2018office, mcgill2019challenges}.

An alternative display option is extending information along with the depth of the screen, extruding it into 3D (see Figure \ref{fig:id}, a).  This extrusion might be specifically suitable for additional information that is directly related to the document displayed on the physical screen. This information is often semantically different and should ideally be displayed separately from the main screen but with a spatial relation (specifically with corresponding $x$- and $y$-coordinates) to the document. For example, when reviewing a document, a common task consists of inserting comments. However, there is no natural space for the added comments within the document, and adding them in place can disrupt the layout of the document. Adding the comments in a separate layer, hovering in front of the main document, maintains the contextual relevance while enabling the user to focus on the layer of interest. 



\subsection{Input-Output Mapping}

The touchscreen allows one layer of information at a time to align with the surface of the screen. This layer enjoys the easiest and most accurate input, as the screen supports the user's hand and enables the highest positional sensing, and, sometimes, pressure sensing. Next, we discuss, how to access layers of information that break out of the bounds of the physical screen. 



\subsubsection{Around the Screen Interaction}

One challenge when interacting across screens is how to support both fine-grained selection on an active screen as well as efficient switching between screens. 
Naive implementations that use a fixed control-to-display gain (CD gain), such as stitching or pointer warping for bridging the space between displays \cite{nacenta2008targeting, waldner2010bridging, xiao2011ubiquitous}, do not scale well to large displays due to the required trade-off between high precision pointing in a given region of interest on a display and fast switching between multiple regions of interest \cite{nancel2013high}. Hence, besides raycasting, prior work has proposed multiple strategies for controlling a large output space using a small input space on a touchscreen (e.g.,~\cite{forlines2006hybridpointing, mccallum2009arc, nancel2011precision, nancel2013high}). 

Inspired by those techniques, we propose using a \textit{bimanual selection technique} for allowing both precise input and efficient switching of the active window. This way, the user can move a cursor and select or deselect items inside the active window by simply using the tablet as a touchpad with a suitable CD gain for single screen interaction, see Figure \ref{fig:id}, b. If the user touches the bezel of the touchscreen, which has a width of 2~cm, with their non-dominant hand, a coarse CD gain is activated to enable fluid switching between multiple windows, see Figure \ref{fig:id}, c. 



The CD gain for single screen interaction is a one-to-one mapping of the movement on the touchscreen.
The CD gain for switching between multiple windows is set in such a way that the user has to move their finger 2~cm to switch to the next screen. We evaluated this CD gain along with values of 1~cm and 3~cm in an informal user study with five participants and found that 2~cm was the fastest and preferred by three participants while the other two CD gains were only preferred by one participant each.

As VR headsets also support head-pointing and partially eye-gaze tracking we implemented a second technique: \emph{combined gaze and touch interaction}. Inspired by previous work on combining gaze and manual pointing \cite{zhai1999manual, pfeuffer2015gaze}, we devised a technique in which the combined head and eye gaze provides the context for touch interaction, see Figure \ref{fig:id}, d. With this technique regions of interest are discrete (the individual virtual screens). Hence, when a user gazes at a specific window, the cursor is transferred to that new screen, maintaining the same absolute position in the new screen as in the old screen (i.e.,~the same $(x,y)$ coordinate).

In order to avoid unwanted cursor movements it is possible to use temporal thresholding (e.g., 200-300 ms \cite{zhai1999manual, pfeuffer2015gaze}), although we did not use this in our evaluation in order to allow for faster task completion times. To allow  selection of items at the screen boundary, we use spatial thresholding with an empirically determined threshold of 5\% of the window size in all directions. Specifically, to switch from one screen to another, the user first has to move their eyes beyond the second display boundary by that threshold. While we did not evaluate this threshold parameter in a formal user study, internal testing revealed that this threshold circumvents unwanted display switches and still allows for comfortable switching between display boundaries.

For both techniques, interacting with the now active virtual display can either 1) happen at the original location of the display, essentially turning the touchscreen into an indirect input device; or 2) occur by aligning the virtual display with the physical touchscreen. For example, dragging the finger on the bottom bezel of the touchscreen (or using a two-finger swipe) can rotate virtual windows around the $x$-axis of a scaffolding geometry (either display-by-display in discrete steps or continuously). 

\subsubsection{Depth Interaction}
\label{sec:depthinteraction}

Accessing virtual displays along the depth dimension of the touchscreen can happen either in front, on, or behind the physical screen (see Figure \ref{fig:id}, a).
For virtual displays in front of the screen, direct mid-air selection might be used, but this may cause occlusion of content at layers further away and hinder interaction on the physical screen \cite{bruder2013touch}. Accessing content behind the screen has to take into account the physical presence of the touchscreen, preventing direct access using virtual hand techniques. Users would either need to grab behind the screen using a virtual hand or use raycasting techniques that can handle object occlusion (e.g., \cite{feiner2003flexible}). Instead, we propose to use the touchscreen to select a desired virtual display. Depending on the number of virtual displays spread behind or in front of the physical screen, different mappings can be appropriate. We experimented with swiping along the bezel of the screen and two-finger swipe gestures, commonly available on multi-touch trackpads.  Depending on the number of layers an absolute mapping for quick random access to layers or a relative mapping for navigating between adjacent layers might be appropriate. In an informal user study with three users, we found that swiping with two fingers outperformed bezel swipes and a relative mapping outperformed an absolute mapping for four and ten layers. Hence, we used this input mapping for the further development of the applications. We used the same CD gain as for switching between individual screens spread around the active screen, as described above.


While the input technique allows to quickly switch between layers it does not completely resolve potential occlusion issues. Occlusion can be mitigated (but not completely resolved) utilizing parallax if the user moves the head. We experimented with two visualization techniques to further mitigate potential occlusion. One option is to amplify the parallax effect when moving the head to the side of the screen, see Figure \ref{fig:retargeting}, f. Another option, inspired by explosion diagrams, is to temporarily scatter virtual screens around the physical screen, see Figure \ref{fig:retargeting}, e. While the explosion diagram technique allows the user to view all layers simultaneously and unoccluded, the number of concurrently visible layers is constrained by the available screen space. Also for both techniques, potential dependencies between layers are impacted (e.g., when there is a need for carefully arranging multiple graphical elements into a composite image). Another limitation of the explosion diagram technique is the reduction of the ability to associate different objects between different layers based on their common 2D locations.

\subsubsection{Single Screen Interaction}
\label{sec:retargeting}
Even an individual layer that is aligned with the physical screen does not need to retain its physical bounds. The ability to display the view frustum according to the rotation of the user's head enables a virtual display that potentially spreads over a very large angular part of the user's field of view. It enables the simulation of screens that may be bigger than physical screens available to the user (see Figure \ref{fig:retargeting}, d). In this case, the mapping of the input space of the physical touchscreen can become indirect. For example, it is possible to use a different CD gain for absolute mapping of the physical finger position to the virtual finger position (following ideas from haptic retargeting \cite{azmandian2016haptic, cheng2017sparse}), see Figure \ref{fig:retargeting}, or the touchscreen can be operated using a relative mapping. 
In some cases, it might be desirable to retain a direct mapping between the input and the output space. In this case, only the portion co-located with the physical touchscreen allows interactivity. Changing this active input area can either be realized by two-finger swipes or input on the bezels of the screen (analogous to implementations in desktop sharing applications, such as TeamViewer) or by redefining the active area by gaze (again requiring an appropriate clutch mechanism). \


\section{Evaluation of Design Parameters}

While we designed the techniques following an iterative approach with multiple design iterations consisting of conceptualization, implementation, and initial user tests (eating your own dog food) \cite{unger2012project, drachen2018games}, we aimed to understand properties of the proposed interaction space in this paper in more detail. Within the proposed interactive space, users can both extend the display area around the current display on a two-dimensional proxy geometry as well as in-depth in front or behind the physical touch screen. For the evaluation, we investigated these two properties (around and in-depth) separately. 

To this end, we first aimed at exploring, if using combined touch and gaze interaction has benefits over touch-only interaction when interacting across multiple virtual screens arranged on a proxy geometry around the touchscreen. Second, we wanted to quantify the benefits of viewing multiple stacked information layers in-depth behind a touchscreen compared to only showing a single layer at a time as is common in many applications today. While adding depth cues has been indicated to improve task performance in various settings \cite{sollenberger1993effects, barfield1999effects, raja2004exploring, ragan2012studying}, we aimed at quantifying any performance improvement within the scope of the joint tablet-HMD interaction.

Hence, we investigated those two aspects in a user study. In the first part (subsequently called the \textsc{Content Transfer Task}), we compared the performance of the \textit{bimanual selection technique} with the \textsc{combined gaze and touch interaction} technique in a content transfer task for a small and large number of screens.

In the second part (subsequently called the \textsc{Puzzle Task}), participants were asked to solve a puzzle task where each puzzle piece was displayed in an individual layer, mimicking tasks when composing presentations using multiple shapes or images using multiple layers in an image editing application. Note that we used the baseline technique described in Section \textit{Depth Interaction} without amplified parallax or explosion diagram visualization due to the nature of the task having strong spatial dependencies between the $x$- and $y$ position of the individual puzzle pieces.




\begin{figure}[t]
	\centering 
	\includegraphics[width=1.0\columnwidth]{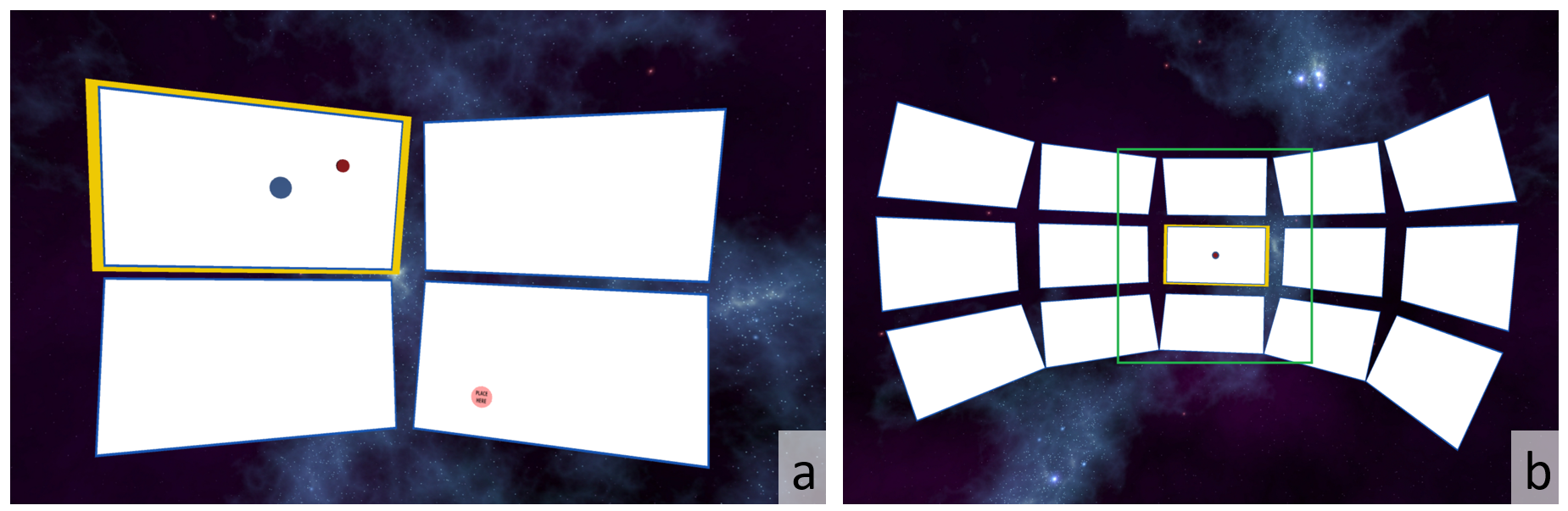}
	\caption{Arrangements of the virtual screens for the \textsc{content transfer task}. a: \textsc{four screens}: the blue dot can be acquired by the red cursor in the upper left screen and is to be placed on the goal in the lower right screen. b: \textsc{fifteen screens}: the green border indicates the field of view of the user. }
	\label{fig:2Dconditions}
\end{figure}
  

For the \textsc{content transfer task}, we followed a $2 \times 2$ within-subjects design. The first independent variable was  \textsc{Interaction Technique} for selecting the screens, which was done either \textsc{Gaze-based} or \textsc{Bimanual} as described above. The second independent variable was  \textsc{Number of Screens} presented to the participant which was either \textsc{Four Screens} (Figure \ref{fig:2Dconditions}, a) or \textsc{Fifteen Screens} (Figure \ref{fig:2Dconditions}, b). We chose these configurations to better understand the performance of the techniques in presence of a few or many screens while still considering the field of view that can be comfortably covered when the user turns their head. This experimental design results in four different conditions, which are depicted in Figure \ref{fig:2Dconditions}.  Dependent variables included tasks completion time (the duration from acquiring the content in one screen to placing it at the marked spot in another screen), accuracy (measured as Euclidean distance between the released item and the actual target location), usability as measured by the System Usability Scale (SUS) \cite{brooke1996sus}, workload as measured by NASA TLX (unweighted version) \cite{hart1988development}, simulator sickness (SSQ) \cite{kennedy1993simulator} as well as user preferences.  We hypothesized that the \textsc{Gaze-based} would outperform \textsc{Bimanual} in terms of task completion time, usability and workload, but not in terms of simulator sickness or accuracy (as the final movements were conducted with the finger of the dominant hand in both techniques). 
 
\begin{figure}[t]
	\centering 
	\includegraphics[width=1.0\columnwidth]{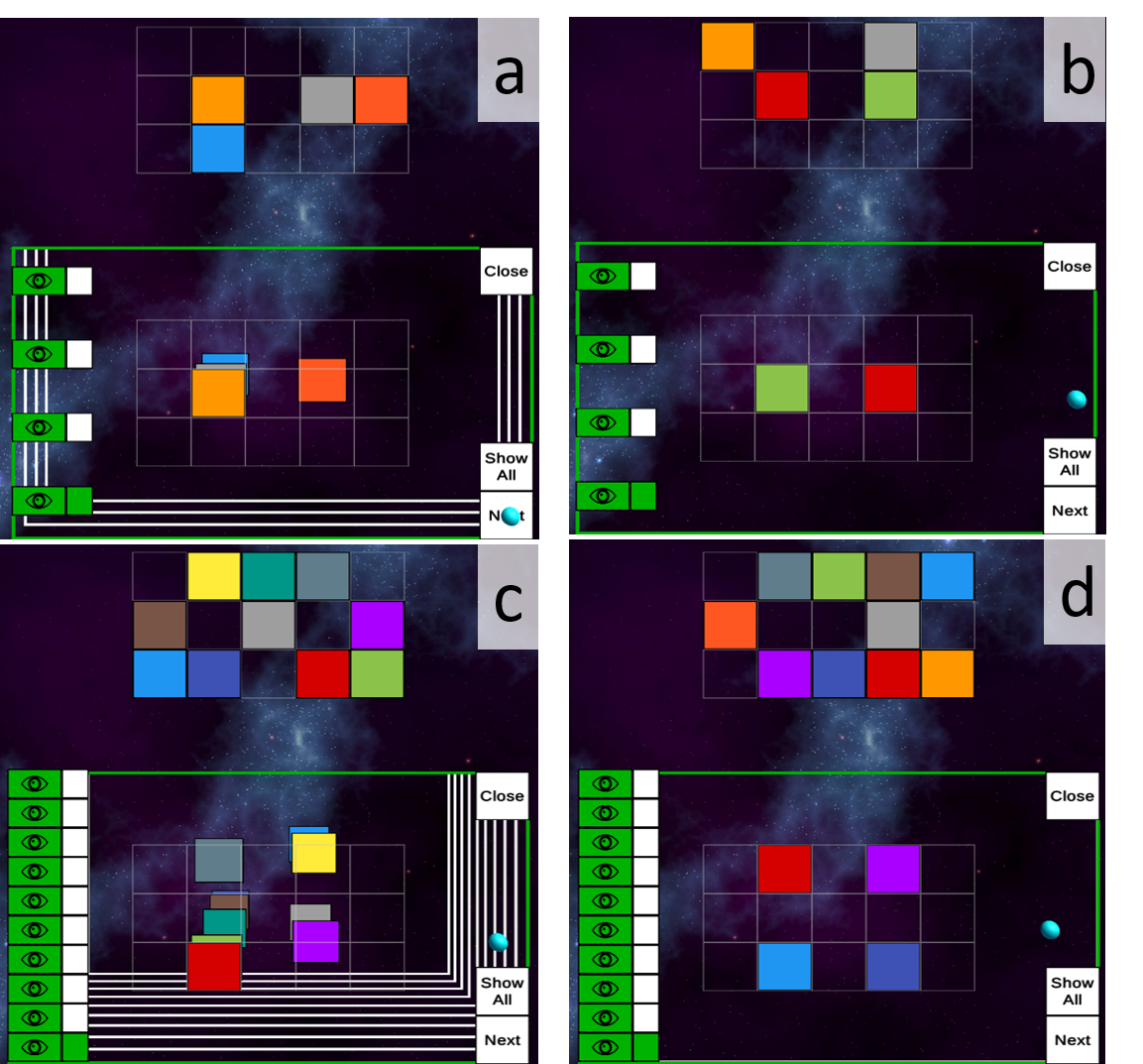}
	\caption{Conditions for the \textsc{Puzzle Task}. a: \textsc{Depth} with \textsc{Four Layers}, b: \textsc{Flat} with \textsc{Four Layers}, c: \textsc{Depth} with \textsc{Ten Layers}, d: \textsc{Flat} with \textsc{Ten Layers}. } 
	\label{fig:conditionspuzzle}
	\vspace{-0.5cm}
\end{figure}

For the \textsc{Puzzle Task}, we also followed a $2 \times 2$ within subjects design. The independent variables were \textsc{Visualization} with two levels: \textsc{Flat}, a baseline condition in which all layers were displayed at the same depth, see Figure \ref{fig:conditionspuzzle}, b and d, and \textsc{Depth}, where each layer was displayed with increasing $z$-distance, see Figure \ref{fig:conditionspuzzle}, a and c. The second independent variable was \textsc{Number of Layers} with two levels: \textsc{Four Layers} with four layers displayed (Figure \ref{fig:conditionspuzzle}, a) and \textsc{Ten Layers} in which ten layers were displayed (Figure \ref{fig:conditionspuzzle}, c). This was done to investigate if an increasing number of layers has an effect on the performance of the visualization techniques.
   
As can be seen in Figure~\ref{fig:conditionspuzzle}, the active layer is highlighted with a green frame, the user's fingertip is visualized with a turquoise sphere. The target arrangement is shown above the puzzle's layered display as a flat image. To the left of the layered display there is an overview widget, inspired by similar views in image editing applications. The eye symbol toggles the visibility of the respective layer (green: visible) and the rectangles to the right of the eye symbol allow direct selection of a layer (green: active layer, white: inactive layer). By selecting a layer (pressing on the green square associated with it), the system moved this layer to align with the screen depth and made all layers in front of it transparent, thereby maximizing the current layer visibility. Switching between two adjacent layers can also be achieved by swiping vertically with two fingers on the touchpad. The button ``Show All'' collapses all layers into a single layer and activates the visibility of all layers. The button ``Next'' switches to the next task. The button ``Close'' was only visible during training and allowed participants to end the training session when they felt comfortable with conducting the task.
   
The dependent variables were, as in the \textsc{Content Transfer task}, task completion time, usability (SUS), workload (TLX) as well as simulator sickness (SSQ). In addition, the system logged the number of incorrectly placed puzzle items.  While we hypothesized that \textsc{Depth} would outperform \textsc{Flat} in terms of task completion time due to the added depth cues, we were specifically interested in quantifying this difference. We also hypothesized that \textsc{Depth} would lead to significantly higher usability rating and lower workload compared to \textsc{Flat} with no difference in simulator sickness. We still included SSQ to check if severe symptoms would occur.

\subsection{Participants}
We recruited 14 participants (5 female, 9 male, mean age 30.07 years, sd = 10.59, mean height 176.57~cm, sd = 8.54).
All of them indicated prior VR experience. Four participants used head mounted VR devices very frequently, two often, four sometimes, three rarely and one only once. 
One participant indicated she does not play video games, two rarely, four participants sometimes, three participants often and four participants very frequently. 
Seven participants wore contact lenses or glasses with corrected to normal vision. 
Eleven participants were right handed while three were left handed. 
All but one used their dominant hand to control the cursor and their non-dominant hand on the bezel. 
Thirteen participants used their index finger on the touchscreen and one used her middle finger.

\subsection{Apparatus}
 The experiment took place in two locations: a laboratory environment in Europe and a home in the US. The study setup (except the PC) was replicated in both environments. A HTC Vive Pro Eye was used as VR HMD with built-in eyetracking. For touch input a Huawei Media Pad T5 was used (screen diagonal 10.1 inches, 16:10 aspect ratio). Velcro was attached to the left, right and bottom bezel of the tablet to support participants in identifying the touchscreen boundaries. The tablet was placed on a table in front of the participant in such a way that participants sitting in front of it on a chair could comfortably use it. The system was implemented in Unity 2019.3 and deployed on a PC with windows 10, an AMD Ryzen Threadripper 1950X 16-core processor with 3.4 Ghz, two Nvidia GeForce RTX 2080 graphics cards and 64 GB RAM in Europe and on a PC with  Intel Core i9-9980HK 8-core processor, Nvidia GeForce RTX 2080 graphics card, 16 GB RAM in the US.
 
\begin{figure}[t]
	\centering 
	\includegraphics[width=0.8\columnwidth]{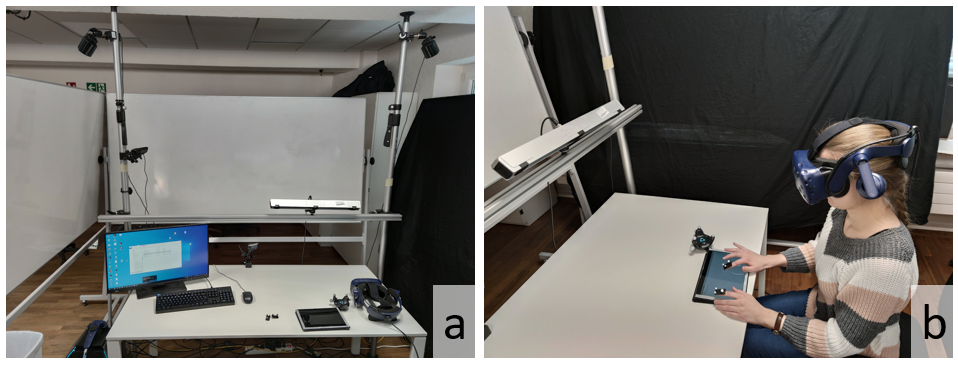}
	\caption{The study setup.  a: setup with motion tracked tablet, fiducials for the fingers and HTC Vive Pro Eye to the right. The lighthouse systems were placed on the ceiling around the table and the Optitrack system on a bar above the table. The monitor, keyboard and mouse to the left were operated by the experimenter to start individual conditions. b: Participant during the study.}
	\label{fig:setup}
	\vspace{-0.7cm}
\end{figure}

 An OptiTrack V120 Trio tracking system was used for spatial tracking of index finger tips by using two rigid bodies with retro-reflective markers attached to them. Note that motion trajectories of the fingers were logged for comparative analysis in future experiments only, they were not analyzed within the scope of this paper. 
 The Vive Lighthouse tracking system was used to track the VR HMD and the tablet. However, while the position of the tablet could have been changed by the participants during the experiment, no participant chose to do so. The study setup is shown in Figure \ref{fig:setup}.
 
 For the \textsc{Content Transfer Task}, multiple screens were placed around the participant's head as shown in Figure \ref{fig:2Dconditions}, a and b. The virtual screen size was 24 inches and virtual screens were initially placed at a distance of 90 cm around the participant's head (if the participant moved their head forwards or backwards this distance changed). The horizontal angle spanned by the screens was at most 190 degrees and the vertical angle less than 30 degrees upwards and less than 35 degrees downwards. This allowed the participant to comfortably look at all screens. The cursor position was indicated as a disc with a diameter of 2.5 cm. For the present study, we chose to use no temporal threshold in order to obtain a comparable switching effect between both techniques.
 
 The \textsc{Puzzle Task} used the same screen size and the same distance to the participant. The individual layers were placed 2.5~cm apart (this parameter was determined empirically). 
 
\subsection{Tasks}
 For the \textsc{Content Transfer Task} several windows were placed in a circle around the participant's head. Then they had to select a disk with a diameter of about 4 cm on one screen to a target area with the same diameter on another screen as shown in \hl{Figure} \ref{fig:2Dconditions} a. The window with the next target area was randomly selected while making sure that both small and longer distances were included during the condition. The disk and the target area appeared at one of eight randomly selected places around the center of a window. Both \textsc{interaction technique}s used the tablet as a touch pad to control the cursor within the active screen and to grab and release the dot via long-press. After successful placement of the disk onto the target area (with at least a partial overlap of the two), the task was repeated with a different arrangement of start and target areas. 
 
  For the \textsc{puzzle task}, participants were asked to arrange puzzle pieces ($5 \times 5$~cm) as indicated by a template image, see Figure \ref{fig:conditionspuzzle}. Each puzzle piece was placed on a separate layer. The puzzle pieces snapped into a predefined  $5 \times 3$ grid to facilitate accurate placement. 
 
  Participants were asked to advance to the next puzzle if they felt they had completed the task. The task was then repeated with a different template.

\subsection{Procedure}

After an introduction, participants were first asked to fill out a demographic questionnaire. Then the HTC Vive Pro Eye eye-calibration procedure was performed to ensure a working eye-tracking calibration for each participant. Thereafter the participants either started with the \textsc{Content Transfer Task} or the  \textsc{Puzzle Task} (counterbalanced). For both tasks, the conditions were ordered using a balanced Latin square. \hl{This resulted in four permutations which could not be equally distributed among 14 participants. However, no significant order effects were detected.} For each condition in the \textsc{Content Transfer Task}, participants performed a training phase where they completed ten content transfers. After that, they performed the actual task 32 times. The number of repetitions was chosen so that each distance and orientation in the \textsc{Fifteen Screens} condition appeared once (1--4 screens horizontally $\times$ \hl{1--2} screens vertically $\times$ two directions each = 8 $\times$ 2 $\times$ 2 = 32). 
For each condition in the \textsc{Puzzle Task}, participants  also performed a training phase where they completed puzzle tasks until they felt comfortable (most users only conducted one practice task). Thereafter, they performed the actual puzzle task ten times per condition. 
In both tasks, after each condition,  participants completed the SSQ, SUS and \hl{NASA} TLX questionnaires. At the end of each block they also answered a questionnaire about their preferences and participated in a semi-structured interview. 
Finally, the participants were thanked and compensated with voucher worth 10 Euro.

\hl{For the \textsc{Content Transfer Task}, we collected 32 repetitions $\times$ 4 conditions $\times$ 14 users = 1792 data points for the performance data and 4 conditions $\times$ 14 users = 56 data points for the subjective feedback.
For the \textsc{Puzzle Task}, we collected 10 repetitions $\times$ 4 conditions $\times$ 14 users = 560 data points for the performance data and 4 conditions $\times$ 14 users = 56 data points for the subjective feedback.}


\subsection{Results}

Statistical significance tests for log-transformed target acquisition time was carried out using general linear model repeated measures analysis of variance (RM-ANOVA) with Holm-Bonferroni adjustments for multiple comparisons at an initial significance level $\alpha = 0.05$. 
We indicate effect sizes whenever feasible ($\eta^2_p$). 


For subjective feedback, or data that did not follow a normal distribution or could not be transformed to a normal distribution using the log-transform (such as errors), we employed the Aligned Rank Transform \cite{wobbrock2011aligned} before applying RM-ANOVAs.

The analysis of results did not reveal significant differences between the participant pool in the US and the one in Europe. Hence, we will solely report the joint results.

Due to logging errors in the \textsc{Content Transfer Task}, we only obtained data from 31 instead of 32 tasks. For one  participant we only received about half of the data in one \textsc{Gaze-based} condition due to technical problems. Four values for the distance between goal and disc were excluded, because they were much higher than should be possible when successfully placing the disc. Note that slightly fewer values for some participants should not affect the overall results as we always used the mean performance for each participant in each condition.

The results in the following sections can be summarized as follows: 
For the \textsc{Content Transfer Task}, participants acquired targets significantly faster with the \textsc{Gaze-based} method compared to \textsc{Bimanual} (ca. 30\%) and the \textsc{gaze-based} resulted in significantly higher SUS ratings. No significant differences regarding accuracy, number of errors or simulator sickness ratings were detected between conditions. All but one participant preferred the \textsc{gaze-based} technique.

For the \textsc{Puzzle Task}, participants performed the task significantly faster with \textsc{Depth} visualization compared to \textsc{Flat} (approximately 15 \%) but also made significantly more errors. The \textsc{Depth} visualization resulted in significantly lower mental demand and resulted in a significantly higher usability rating compared to \textsc{Flat}. 


\subsubsection{Performance} 
\begin{figure*}[t]
	\centering 
	\includegraphics[width=2.05\columnwidth]{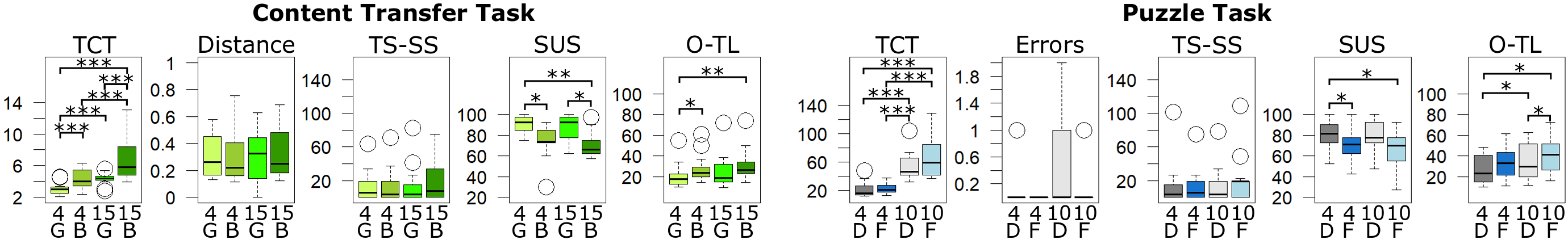}
	\caption{Results of the task completion time (TCT), the Total Severity aspect of the Simulator Sickenss Questionnaire (TS-SS), the System Usability Scale (SUS) and the overall taskload (O-TL) for the \textsc{Content Transfer Task} and the \textsc{Puzzle Task}. Also for the \textsc{Content Transfer Task} the distance between disc and goal in cm upon task completion is shown and for the \textsc{Puzzle Task} the mean number of total errors. On the $x$-axis the digits indicates the \textsc{Number of Screens} (4 or 15) or the \textsc{Number of Layers} (4 or 10) as well as the \textsc{Interaction Technique} G = \textsc{Gaze-based}, B = \textsc{Bimanual} and the \textsc{Visualization} F = \textsc{Flat}, D = \textsc{Depth}.
	The number of stars indicates the level of significance between the conditions (*** $<0.001$ ** $<0.01$ * $<0.05$).}
	\label{fig:all_plots}
	\vspace{-0.6cm}
\end{figure*}

For the \textsc{Content Transfer Task} there was a significant main effect of \textsc{Interaction Technique} for task completion time, such that the \textsc{Gaze-based} method ($M=3.70~s, SD=0.98$) resulted in a shorter task completion time than the \textsc{Bimanual} method ($M=5.49~s, SD=2.33$). As expected, the main effect of the \textsc{Number of Screens} on the task completion time was also significant such that \textsc{Four Screens} ($M=3.66~s, SD=1.20$) resulted in a shorter task completion time than \textsc{Fifteen Screens} ($M=5.52~s, SD=2.20$). This was predictable because moving the target across multiple columns and rows takes longer than moving it only across two columns and rows. There was no significant interaction effect between \textsc{Interaction Technique} and \textsc{Number of Screens} with respect to task completion time. 
There was no significant differences between the conditions for accuracy in placing the target on the goal. The performance results for the \textsc{Content Transfer Task} can be seen in Figure \ref{fig:all_plots} and the results of the RM-ANOVA in Table \ref{tab:PerformanceResults}.

For the \textsc{Puzzle Task} there was a significant main effect of \textsc{Visualization} for task completion time such that the \textsc{Depth} method ($M=37.17~s, SD=22.79$) resulted in a shorter task completion time than the \textsc{Flat} method ($M=43.31~s, SD=29.52$). As expected, the main effect of \textsc{Number of Layers} for task completion time was significant such that the \textsc{Four Layers} ($M=21.28~s, SD=8.87$) resulted in a shorter task completion time than \textsc{Ten Layers} ($M=59.19~s, SD=24.23$). No significant interactions have been found between \textsc{Number of Layers} and \textsc{Visualization}.
Analysis of the error data (using the total number of errors across all ten repetitions) indicated that both the \textsc{Number of Layers} and the \textsc{Visualization} method had a significant impact on the number of errors made.
The conditions using the \textsc{Depth} method ($M=0.29$, $SD=0.6$) had significantly more errors than the conditions using the \textsc{Flat} method ($M=0.07$, $0.26$). Furthermore, conditions with \textsc{Four Layers} ($M=0.04$, $SD=0.19$) resulted in significantly fewer errors than conditions with \textsc{Ten Layers} ($M=0.32$, $0.61$). 


\begin{table*}[t]
\centering 
\tiny
\begingroup
\setlength{\tabcolsep}{5pt}
\begin{tabular}{|>{\centering}m{0.38cm}||>{\centering}m{0.2cm}|>{\centering}m{0.2cm}|c|c|c||>{\centering}m{0.2cm}|>{\centering}m{0.2cm}|c|c|c||>{\centering}m{0.2cm}|>{\centering}m{0.2cm}|c|c|c||c|c|c|c|c||>{\centering}m{0.2cm}|>{\centering}m{0.2cm}|c|c|c|}
\multicolumn{26}{c}{\small\bfseries\textbf{Content Transfer Task}} \\
\hline 
\multicolumn{6}{|c||}{Task Completion Time} & \multicolumn{5}{c||}{Accuracy} & \multicolumn{5}{c||}{TS-SS} & \multicolumn{5}{c||}{SUS} & \multicolumn{5}{c|}{Overall Taskload}\\
\hline 
 & d$f_{1}$ & d$f_{2}$ & F & p &  $\eta^2_p$ & d$f_{1}$ & d$f_{2}$ & F & p &  $\eta^2_p$ & d$f_{1}$ & d$f_{2}$ & F & p &  $\eta^2_p$ & d$f_{1}$ & d$f_{2}$ & F & p &  $\eta^2_p$ & d$f_{1}$ & d$f_{2}$ & F & p &  $\eta^2_p$\\ 
\hline 

I & \cellcolor{lightgray}$\mathbf{1}$ & \cellcolor{lightgray}$\mathbf{13}$ & \cellcolor{lightgray}$\mathbf{63.66}$ & \cellcolor{lightgray}$\mathbf{<.001}$ & \cellcolor{lightgray}$\mathbf{70.83}$ &  $1$ & $13$ & $.79$ & $.39$ & $.06$ &  \cellcolor{lightgray}$\mathbf{1}$ & \cellcolor{lightgray}$\mathbf{13}$ & \cellcolor{lightgray}$\mathbf{11.11}$ & \cellcolor{lightgray}$\mathbf{.005}$ & \cellcolor{lightgray}$\mathbf{.46}$ &  \cellcolor{lightgray}$\mathbf{1}$ & \cellcolor{lightgray}$\mathbf{13}$ & \cellcolor{lightgray}$\mathbf{38.42}$ & \cellcolor{lightgray}$\mathbf{<.001}$ & \cellcolor{lightgray}$\mathbf{.75}$ &  \cellcolor{lightgray}$\mathbf{1}$ & \cellcolor{lightgray}$\mathbf{13}$ & \cellcolor{lightgray}$\mathbf{29.81}$ & \cellcolor{lightgray}$\mathbf{<.001}$ & \cellcolor{lightgray}$\mathbf{.7}$\\ 

S & \cellcolor{lightgray}$\mathbf{1}$ & \cellcolor{lightgray}$\mathbf{13}$ & \cellcolor{lightgray}$\mathbf{99.41}$ & \cellcolor{lightgray}$\mathbf{<.001}$ & \cellcolor{lightgray}$\mathbf{.88}$ &  $1$ & $13$ & $.53$ & $.48$ & $.04$ &  $1$ & $13$ & $.60$ & $.45$ & $.04$ &  $1$ & $13$ & $3.79$ & $.07$ & $.23$ &  $1$ & $13$ & $3.02$ & $.11$ & $.19$\\ 

I $\times$ S & $1$ & $13$ & $.86$ & $.37$ & $.06$  & $1$ & $13$ & $1.2$ & $.29$ & $.08$ &  $1$ & $13$ & $2.0$ & $.18$ & $.13$ &  $1$ & $13$ & $.87$ & $.37$ & $.06$ &  $1$ & $13$ & $.14$ & $.72$ & $.01$\\ 
\hline 
\end{tabular}

\begin{tabular}{|>{\centering}m{0.48cm}||>{\centering}m{0.2cm}|>{\centering}m{0.2cm}|c|c|c||>{\centering}m{0.2cm}|>{\centering}m{0.2cm}|c|c|c||>{\centering}m{0.2cm}|>{\centering}m{0.2cm}|c|c|c||>{\centering}m{0.2cm}|>{\centering}m{0.2cm}|c|c|c||>{\centering}m{0.2cm}|>{\centering}m{0.2cm}|c|c|c|}
\multicolumn{26}{c}{\small\bfseries\textbf{Puzzle Task}} \\
\hline
\multicolumn{6}{|c||}{Task Completion Time} &\multicolumn{5}{c||}{Errors} & \multicolumn{5}{c||}{TS-SS}  & \multicolumn{5}{c||}{SUS} & \multicolumn{5}{c|}{Overall Taskload} \\
\hline 
 & d$f_{1}$ & d$f_{2}$ & F & p &  $\eta^2_p$   & d$f_{1}$ & d$f_{2}$ & F & p &  $\eta^2_p$ & d$f_{1}$ & d$f_{2}$ & F & p &  $\eta^2_p$ & d$f_{1}$ & d$f_{2}$ & F & p &  $\eta^2_p$ & d$f_{1}$ & d$f_{2}$ & F & p &  $\eta^2_p$\\ 
\hline 
V & \cellcolor{lightgray}$\mathbf{1}$ & \cellcolor{lightgray}$\mathbf{13}$ & \cellcolor{lightgray}$\mathbf{5.32}$ & \cellcolor{lightgray}$\mathbf{.04}$ & \cellcolor{lightgray}$\mathbf{.29}$  & \cellcolor{lightgray}$\mathbf{1}$ & \cellcolor{lightgray}$\mathbf{13}$ & \cellcolor{lightgray}$\mathbf{8.97}$ & \cellcolor{lightgray}$\mathbf{.04}$ & \cellcolor{lightgray}$\mathbf{.28}$ &  \cellcolor{lightgray}$\mathbf{1}$ & \cellcolor{lightgray}$\mathbf{13}$ & \cellcolor{lightgray}$\mathbf{12.45}$ & \cellcolor{lightgray}$\mathbf{.004}$ & \cellcolor{lightgray}$\mathbf{.49}$ &  \cellcolor{lightgray}$\mathbf{1}$ & \cellcolor{lightgray}$\mathbf{13}$ & \cellcolor{lightgray}$\mathbf{10.39}$ & \cellcolor{lightgray}$\mathbf{.007}$ & \cellcolor{lightgray}$\mathbf{.44}$ &  \cellcolor{lightgray}$\mathbf{1}$ & \cellcolor{lightgray}$\mathbf{13}$ & \cellcolor{lightgray}$\mathbf{12.34}$ & \cellcolor{lightgray}$\mathbf{.004}$ & \cellcolor{lightgray}$\mathbf{.49}$\\ 

L & \cellcolor{lightgray}$\mathbf{1}$ & \cellcolor{lightgray}$\mathbf{13}$ & \cellcolor{lightgray}$\mathbf{1574.78}$ & \cellcolor{lightgray}$\mathbf{<.001}$ & \cellcolor{lightgray}$\mathbf{.99}$  & \cellcolor{lightgray}$\mathbf{1}$ & \cellcolor{lightgray}$\mathbf{13}$ & \cellcolor{lightgray}$\mathbf{5.14}$ & \cellcolor{lightgray}$\mathbf{.04}$ & \cellcolor{lightgray}$\mathbf{.28}$ &  $1$ & $13$ & $4.04$ & $.07$ & $.24$ &  $1$ & $13$ & $.61$ & $.45$ & $.04$ &  \cellcolor{lightgray}$\mathbf{1}$ & \cellcolor{lightgray}$\mathbf{13}$ & \cellcolor{lightgray}$\mathbf{15.4}$ & \cellcolor{lightgray}$\mathbf{.002}$ & \cellcolor{lightgray}$\mathbf{.54}$\\ 

V $\times$ L & $1$ & $13$ & $1.23$ & $.29$ & $.09$   & \cellcolor{lightgray}$\mathbf{1}$ & \cellcolor{lightgray}$\mathbf{13}$ & \cellcolor{lightgray}$\mathbf{4.7}$ & \cellcolor{lightgray}$\mathbf{.05}$ & \cellcolor{lightgray}$\mathbf{.03}$ &  $1$ & $13$ & $1.86$ & $.2$ & $.13$ &  $1$ & $13$ & $.08$ & $.78$ & $.0$ &  $1$ & $13$ & $.03$ & $.86$ & $.0$\\ 
\hline 
\end{tabular}
\endgroup

\caption{RM-ANOVA results for both tasks. Gray rows show significant findings. I = \textsc{Interaction Technique}, S = \textsc{Number of Screens}, V = \textsc{Visualization}, L = \textsc{Number of Layers}. TS-SS: Total Severity Dimension of the Simulator Sickness Questionnaire. SUS: System Usability Scale d$f_1$ = d$f_{effect}$ and d$f_2$ = d$f_{error}$.}
\label{tab:PerformanceResults}
\end{table*}



\vspace{-0.2cm}
\subsubsection{Simulator Sickness, Workload, Usability}
\vspace{-0.15cm}
For the \textsc{Content Transfer Task} we found \hl{that the Total Severity aspect of the Simulator Sickness Questionnaire was significantly influenced by the \textsc{Interaction Technique} such that \textsc{bimanual} ($M=14.43$, $SD=20.73$) resulted in a higher total severity than \textsc{gaze-based} ($M=13.35$, $20.57$). Also the Oculo-motor aspect was lower for the \textsc{gaze-based} than for the \textsc{bimanual}. However, the values are very low and pairwise comparisons showed no significant differences.} 
There was a significant difference for the overall TLX results, such that \textsc{Bimanual} resulted in a higher taskload than \textsc{Gaze-based}. 
\hl{Also the mental, physical, effort and frustration results of the TLX were significantly higher for the \textsc{Bimanual} conditions than for the \textsc{gaze-based} conditions.}
The perceived performance of the participants, however, was significantly higher for the \textsc{Bimanual} method ($M=27.14$, $SD=19.31$) compared to the \textsc{Gaze-based} method ($M=21.79$, $SD=21.65$). This is in contrast to the findings of the task completion time and distance to goal shown above, where task completion time was significantly shorter for \textsc{Gaze-based}, and no significant difference was found in distance to the goal.
Furthermore, \textsc{Interaction Technique} had a significant impact on the SUS score in such a way that the usability of the \textsc{Gaze-based} method ($M=88.75, SD=10.26$) was higher than the usability of the \textsc{Bimanual} method ($M=72.59, SD=13.63$). The main results of the three questionnaires are shown in Figure \ref{fig:all_plots}. Plots and detailed results on the further dimensions of simulator sickness and task load can be found in the supplementary material.

For the \textsc{Puzzle Task}, the results of the simulator sickness questionnaire indicated \hl{that \textsc{visualization} had a significant impact on the Total Severity such that it was significantly lower for \textsc{depth} ($M=14.29$, $SD=23.85$) than for \textsc{flat} ($M=16.7$, $SD=24.76$). This was also true for the Oculo-motor and disorientation aspect. Also, the \textsc{number of layers} significantly influenced the disorientation aspect, such that the \textsc{four layer} had a significantly lower value than \textsc{ten layer}. However, the values are again very low and no significant differences were found in pairwise comparisons.} 
\hl{The NASA TLX results indicated that the \textsc{visualization} significantly influenced the overall task load such that it was significantly higher for the \textsc{flat} visualization ($M=38.3$, $SD=16.06$) than for the \textsc{depth} visualization ($M=30.92$, $SD=15.92$). This was also the case for the mental and physical task load, as well as the frustration perceived by the participants.
The perceived performance, however, was significantly higher for \textsc{flat} ($M=29.64$, $SD=21.73$) than for \textsc{depth} ($M=22.86$, $SD=22.17$) which again, is in contrast to the results of the task completion time.  
In addition, we found that the number of layers had a significant impact on the overall task load such that it was lower for \textsc{four layers} ($M=30.83$, $SD=14.67$) than for \textsc{ten layers} ($M=38.39$, $SD=17.17$). The same was observed for frustration, physical task load and the temporal aspect.}    
Also, we found that \textsc{Visualization} had a significant impact on the SUS results, in such a way that \textsc{Depth} ($M=80.09$, $SD=14.05$) resulted in a higher usability than \textsc{Flat} ($M=68.48$, $SD=17.00$).  
The main results of the three questionnaires is shown in Figure \ref{fig:all_plots} and Table \ref{tab:PerformanceResults}. Plots and detailed results on the further dimensions of simulator sickness and task load can be found in the supplementary material.





\subsubsection{Preference, Open Comments and Observations}

For the \textsc{Content Transfer Task}, all but one participant preferred the \textsc{Gaze-based} method over the \textsc{Bimanual}, regardless of the number of screens. However, when asked about their preference for each \textsc{Number of Screens} separately this participant also liked \textsc{Gaze-based} more for \textsc{Fifteen} screens. When asked which interaction method the participants perceived as faster, all participants chose \textsc{Gaze-based}. When asked which method they could interact with more precisely, all participants chose \textsc{Gaze-based} for the conditions with \textsc{Fifteen Screens}. However, in the condition with \textsc{Four Screens}, only four participants thought that \textsc{Gaze-based} allowed for a more precise interaction and three participants felt that \textsc{Bimanual} was more precise. This is also supported by the objective measurements shown in  Figure \ref{fig:all_plots}, even though no significant difference between conditions was indicated for distance to the target. We conjecture that these differences are due to the fact that in the \textsc{gaze-based} method the selection of the target screen, and moving to the target within the screen, \textit{can} happen concurrently. In contrast, in the \textsc{Bimanual} condition the selection of the target screen is \textit{always} separated by an explicit mode switch (raising the finger of the non-dominant hand from the bezel) from the subsequent phase of moving towards the target within that screen. The lower precision in the \textsc{Bimanual} condition with \textsc{Fifteen Screens} compared to \textsc{Four screens} might also be due to higher fatigue when moving the item across more screens. 
The participant who preferred \textsc{Bimanual} explained that it reminds him of the keyboard (alt+tab). Most other participants explained their preference for \textsc{Gaze-based} due to it being faster. One participant found \textsc{Gaze-based} more intuitive, and another participant found it less fatiguing. One participant suggested a hybrid solution where \textsc{Gaze-based} is only enabled when touching the bezel.

For the \textsc{Puzzle Task}, all but one participant preferred the \textsc{Depth} condition. However, one participant preferred \textsc{Flat}. This participant later mentioned having trouble focusing the right eye and therefore found it more difficult to match the 3D state with the 2D target.  Two participants thought the \textsc{Flat} visualization was faster for \textsc{Four Layers} than the \textsc{Depth} visualization. All other participants felt that \textsc{Depth} was faster. Regarding the \textsc{Depth} visualization, participants also mentioned that ``If you make a mistake it is easier to find the layer with the wrong element'' (Participant P01), ``I can better see which puzzle piece is where'' (P02),  ``you directly see where the puzzle piece is'' (P04) and that \textsc{Depth} ``provides more information'' (P06).

Participants used different strategies to solve the puzzle task for the \textsc{Flat} conditions.
Three participants mainly identified the element belonging to the current layer by trial and error. They tried to move the element and if it was not possible they tested the next element until they found the element of that layer. Four participants started using this strategy but then switched to a more efficient one. One participant used it sporadically in the condition with \textsc{Four Layers}.

Some participants moved to the last layer (furthest away) at the start of each puzzle and continued to the front. In the last layer, only one element (the element of that layer) was visible, therefore the participant did not have to guess the correct element. When moving one layer, a new element appeared on the screen, and the participant could identify the next element with ease. Nine participants mainly used this strategy.

There were also participants who started from the front layer but shortly switched to the next layer and back in order to detect the active element as this element will then disappear and reappear. Two participants switched to this strategy after first using another strategy. Also, this strategy was basically always used when participants wanted to correct a mistake.

The fourth strategy observed was to click on the eye-symbols in the menu at the left to deactivate and activate the current layer. This would also show the user which element would disappear and reappear and therefore was the element of the active layer. This strategy was only used by one participant in his first \textsc{Flat} condition. Thereafter he switched strategy. Two people also tried this strategy during the training phase.

\section{Applications}
\label{sec:apps}
Based on the insights presented in the previous sections, we 
implemented six applications that we envision are applicable in mobile knowledge worker scenarios.

\begin{figure*}[htb!]
    \centering 
    \includegraphics[width=1.6\columnwidth]{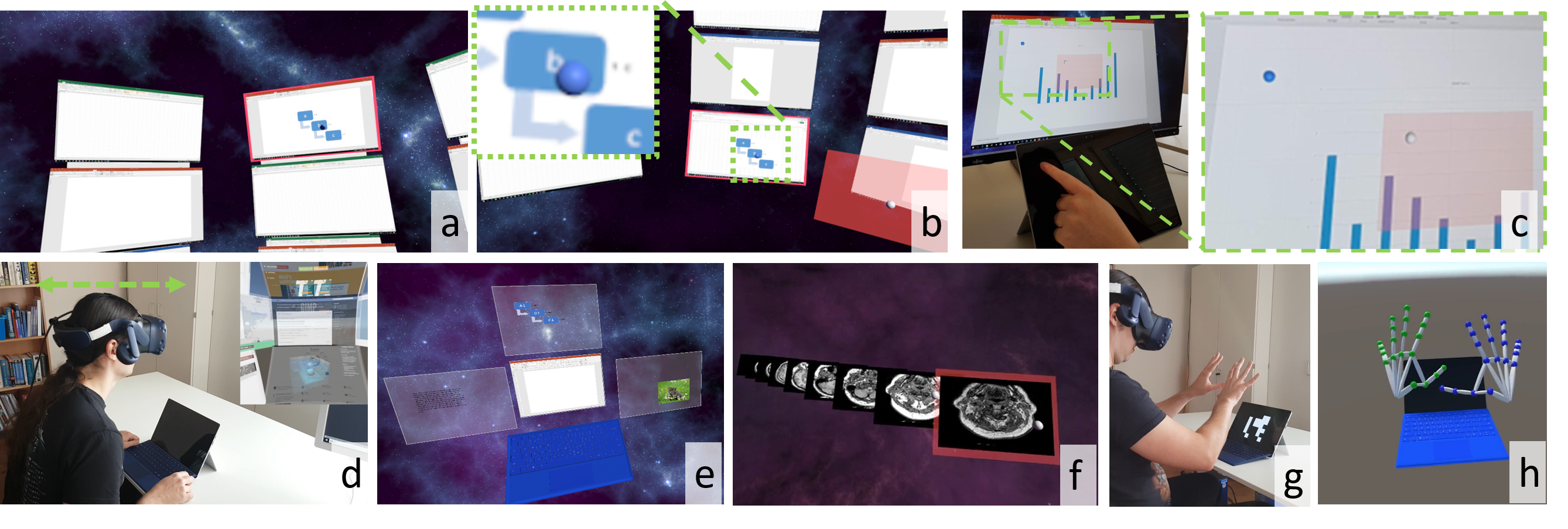}
     \vspace{-0.4cm}
    \caption{a-d: Window Manager. a: An image is selected for movement to another window by head-gaze (selecting the window) + touch (selecting the image inside the window). b: The image is released on a different window at a different layer. The grey sphere on the red rectangle indicates the physical finger position. The blue sphere on the image depicts the projected finger position on the current window. c: The input on and above the physical touchscreen (red transparent rectangle, fingertip on touchscreen indicated by a white sphere) is retargeted to the virtual display (indicated by a blue sphere). d: A user previews  windows in a hidden layer by leaning forward.  e-f: Options to temporarily rearrange information layers. e: Virtual displays are temporarily arranged around the physical screen. f: If a user peeks behind the physical screen, virtual displays are extruded to that side. g-h mobile implementation: g: user wearing an HTC Vive Eye Pro looking on a tablet PC, which displays an Aruco marker. h: the user's fingers are spatially tracked with the HTC Vive Hand tracking SDK}
    \label{fig:retargeting}
    \vspace{-0.6cm}
\end{figure*}

\textbf{Window Manager:} 
We implemented a window manager that allows the arrangement of multiple windows around the user (see Figure \ref{fig:teaser}, b). Interaction with those windows is supported by joint spatial and touch interaction. For example, the active window is selected by head gaze (indicated by a red border around the window in Figure \ref{fig:retargeting} a and b). The input space of the physical touchscreen (depicted as a red rectangle in Figure \ref{fig:retargeting}, b) is then mapped to the output space of the virtual window. Contrary to the content transfer task in the evaluation of design parameters, for the window manager, the selected window stays selected even if the user's gaze is switching to another window for as long as the user is touching the physical touchscreen. This design lets the user glance at other windows without losing the context of the task. Other clutching mechanisms are also feasible, such as locking the gaze-selected window on touch-down but then delaying the release on touch-up to let the user interact with the touchscreen. 

To switch depth layers, the user can either swipe along the bezel or use a two-finger swipe, as described in the Section ``Depth Interaction''. Alternatively, to retrieve a temporary preview of the otherwise hidden layers, the user can lean towards the virtual screen to peak through (Figure \ref{fig:retargeting}, c). For interaction with virtual windows that are larger than the physical touchscreen, retargeting can be used, as described in the Section ``Single Screen Interaction'' (Figure \ref{fig:retargeting}, d).

\textbf{Code Version Control:} We implemented an interface for code version control (Figure \ref{fig:teaser}, c) that uses a spatial layout around the physical screen, see Figure \ref{fig:teaser}, c. Using on-screen pinches the user can select different scopes of code changes (line, block, function, class, file), swipe through the commits using the selected scope, and swipe the desired commit down to the physical screen for further editing.


\textbf{Parallel Coordinates:} Inspired by recent research in immersive analytics \cite{marriott2018immersive, cordeil2017imaxes, tadeja2019exploring, cordeil2020emaxes}, we built a parallel coordinates plot that is grounded on the tangible physical screen, see Figure \ref{fig:teaser}, g, and h. Subranges of variables can be selected using on-surface swipes for the extent of the variable range and on-surface drags for the center of the variable range. Individual variables are selected by touch. Switching between coordinate axes ($z$, protruding the display, for the variable range, $y$, along the height of the tablet, for the data items) can be achieved explicitly using a mode switch (e.g., pressing a soft button) or by mapping the swipe and drag along the $x$- and $y$-axis of the touchscreen.

\textbf{Map:} In the map application, users can navigate on a single layer without the map being cropped at the boundaries of the physical screen, see Figure \ref{fig:teaser}, a . Available alternative map layers can be previewed by tilting the physical screen and be selected by swiping along the bezel of the screen.

\textbf{Medical Imaging:} In the medical imaging application, users can also swipe through different layers of information, see Figure \ref{fig:teaser}, e--f. Additionally, the image slices can be previewed when the user moves their head to the side of the screen to look behind it.

\textbf{PowerPoint:} In the VR PowerPoint application users can arrange graphical items on a canvas, see Figure \ref{fig:teaser}, d. Each item is associated with a separate layer. In the cases when items are occluded, they can be quickly accessed by swiping through the layers. 



\section{Mobile Implementation}

The system was implemented using Unity 2019.3. We used a HTC Vive Pro Eye as the HMD, which also enables hand tracking\footnote{https://developer.vive.com/resources/knowledgebase/vive-hand-tracking-sdk/ Last accessed April 24th, 2020.} as well as access to the camera streams\footnote{https://developer.vive.com/resources/knowledgebase/intro-vive-srworks-sdk/ Last accessed April 24th, 2020.}. While we used external OptiTrack outside-in tracking systems due to superior hand tracking accuracy compared to HMD-based inside out tracking \cite{schneider2020accuracy} for our evaluations, we also implemented a system to track the user's screen using the HMD cameras, see Figure \ref{fig:retargeting}, g-h.  While it is possible to build a computer vision algorithm to track the specific laptop model used by users, we used a model-independent approach. Since the laptop or tablet display screen is hidden from the user, wearing an HMD, we can use the screen to display a standard tracking pattern, such as ARUCO markers \cite{munoz2012aruco}. Displaying the pattern enables a robust detection and orientation using multiple wide-field-of view cameras of the screen and information about the laptop dimension are sufficient to display a virtual model that can guide the user touch gestures (see accompanying video). Other solutions, such as using an external camera, such as a laptop camera, to track the HMD are feasible but require additional instrumentation of the HMD \cite{mohr2019trackcap}.

To facilitate the community in further research and development of joint interactions between touchscreens and VR headsets, we make our code available under https://gitlab.com/mixedrealitylab/breakingthescreen.

\section{User Feedback on Applications}


The purpose of this user study was to learn from initial user reactions on the applications we previously introduced. To this end we demonstrated individual experiences to the user instead of carrying out task-based evaluations (following the approaches used in prior work, e.g., \cite{chen2014duet, schneider2019reconviguration}). We recruited 17 participants (8 female, 9 male, mean age 31.9 years, sd = 9.2, mean height 173.9 cm, sd = 6.8) from a university campus with diverse study backgrounds. All participants were familiar with touch-sensitive screens.

\subsection{Apparatus, Procedure, and Task}


The same apparatus as in the evaluation of the design space parameters were used with the following change: instead of the Android tablet, a motion-tracked Surface Pro 4 tablet was used. 

After an introduction, participants were asked to fill out a demographic questionnaire. Thereafter, we fixated OptiTrack trackers on the index finger of the dominant hand and the thumb of the other hand. Then individual applications were presented (counterbalanced). After trying out the applications the users rated them using a short three-item questionnaire capturing the user experience dimensions of ease of use, utility, and enjoyment (in line with similar approaches in prior work~\cite{lewis1991psychometric, chen2014duet}). After trying out the applications, participants were asked to participate in a semi-structured interview. Finally, the participants were compensated with a 10 Euro voucher.



The order of the applications was balanced across participants, insofar as possible (a full permutation was not possible due to the number of applications).

\subsection{Results}
 \vspace{-0.2cm}

\begin{figure}[t]
	\centering 
	\includegraphics[width=0.9\columnwidth]{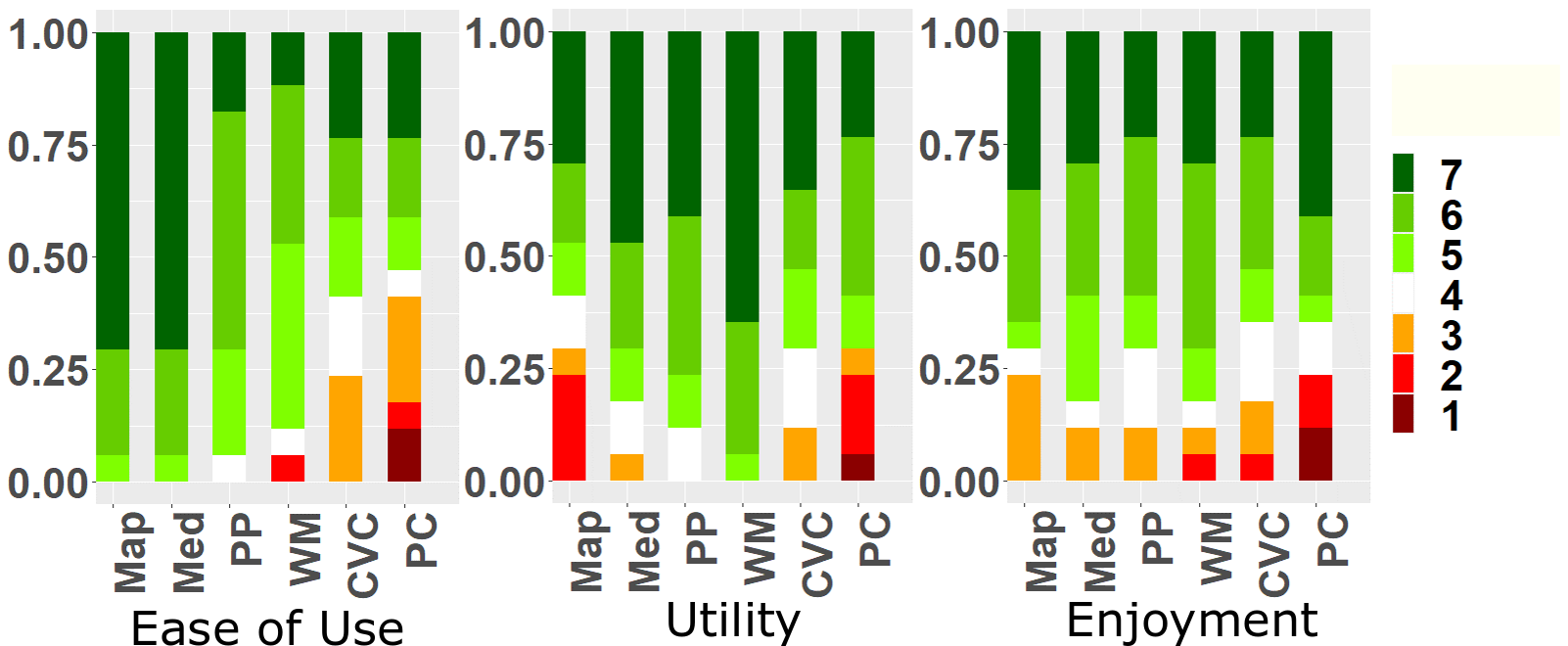}
    \caption{Ease of Use, Utility and Enjoyment ratings for the evaluated applications on a range of 1 (strongly disagree) to 7 (strongly agree). \textsc{Map}: Map Interaction, \textsc{Med}: \textsc{Medical}, \textsc{PP}: \textsc{PowerPoint}, \textsc{WM}: \textsc{WindowManager}, \textsc{CVC}: \textsc{CodeVersionControl}, \textsc{PC}: \textsc{ParallelCoordinates}. The y-axis depcits the number of participants in percent (1.0  = 100\% of participants).}
    \label{fig:appratings}
    \vspace{-0.6cm}
\end{figure}



Figure \ref{fig:appratings} shows user ratings on seven-item Likert scales for questions on ease of use (``I found the application easy to use''), utility (``I found the application to be useful'') and enjoyment (``I had fun interacting with the application''). Note that we did not run null hypothesis significance tests on these ratings as they should serve as a descriptive indication of these user experience dimensions only. Participants were asked to comment on individual applications. We followed top-down qualitative coding and structuring procedures to identify the benefits and drawbacks of the individual applications \cite{strauss1990basics}. For \textsc{WindowManager} the participants appreciated that the need for multiple physical monitors was mitigated (``Avoids the need of a monitor wall.'' participant P13) but also the layered view of the virtual displays (``You can view dozens of documents.'' P03) and it's ease of use (``It was easy to understand even for non-experts.'' P04). Regarding \textsc{CodeVersionControl}, one participant mentioned that she had an ``overview like in a airport control tower.'' P13 and another one saw applicability to another domain: ``I would like that for functional subunits in engineering'' P17. Similarly, for \textsc{PowerPoint} the participants saw a transfer to other domains with one mentioning that ``these layers could also be used in drawing tools like Gimp or Photoshop'' P14. Regarding,  \textsc{ParallelCoordinates} a participant mentioned that ``it is a more entertaining way to present boring data'' P06 and another participant, ``The new visualization was nice'' P13. Regarding \textsc{Map} a participant mentioned that ``I could need that for driving.'' P09 and regarding \textsc{Medical} participants mentioned that ``a 3D Model from the layers would be awesome'' P14 but also that ``It would be cool to pick the layers directly'' P04.

\subsection{Discussion}

Using immersive HMDs enables mobile knowledge workers to bring their working environment with them everywhere they go \cite{grubert2018office}.  Relative to the compact input and output space of today's mobile devices, the large field of view of HMDs allows to substantially extend the available display and increase privacy. \hl{Our design explorations and evaluations, explored different dimensions of this extended display space while focusing the input to the small area of a tablet screen. The first question we investigated was handling the large field of view of the virtual environment while using a very limited input space. To handle the large scale ratio between input and display, we tested two different techniques, One where we combined noisy eye gaze for large scale motions and touch for fine manipulation and compared it with a multi-touch technique.} The results of the design parameter evaluation indicated that combination of gaze and touch input on a tablet as realized in the \textsc{Gaze-based} input method are beneficial for interaction in virtual multi-window environments. Specifically, the \textsc{Gaze-based} method outperformed the \textsc{Bimanual method} by approximately 30\%. 
\hl{The second parameter we studied was the use of the HMD's depth display for interaction above the touch screen. Using multiple layers has benefits of extending the amount of information that can be displayed and manipulated using a 2D screen. On the other-hand depth parallax and occlusions may hinder the user's performance. Our second study was designed to test it, by asking users to do a task both in 2D and in 3D. As indicated by the results, the use of depth can be both efficient and usable when compared to a 2D touch interaction. However, we were surprised by different strategies employed by some users to compensate for the lack of the depth structure in  \textsc{Flat} visualization,} such as ordering their manipulations from back to front. \hl{Pairwise comparisons showed no significant differences between the conditions with regard to errors and} the overall number of errors was low (5.7\% for \textsc{Depth} and 1.4\% for \textsc{Flat}). During interviews with participants, the five participants generating errors stated that they did not use the option to show all layers at the end of the task using the dedicated button shown in Figure \ref{fig:conditionspuzzle}. By design, the layers were arranged behind the physical screen, allowing the user to navigate them front to back. As previously noted, in the \textsc{Flat} visualization nine participants first navigated to the last layer and then navigated to the front layer by layer. When they reached the front layer, all layers with all puzzle pieces were visible, hence there was no need to use the ``Show all'' button. The user feedback on the developed applications indicated that the prototypes were usable and enjoyable. Users could envision employing the prototypes in their work environments. We envision mobile VR to become an important tool for work, as it enables a large and private display on the go, regardless of the worker environment. However the physical environment of the user will not grow and the interactions should be designed accordingly. The presented applications are just a small probe of the complex design space encompassing all of the vast tasks information workers carry out, and this area is ripe for further research and development.

\subsubsection{Use of Augmented Reality HMDs}

The proposed interaction concepts, while being considered in VR, could be transferred to Augmented Reality (AR) and explored further. In AR, current generation optical see-through HMDs typically have a substantially smaller field of view compared to immersive VR HMDs. Hence, the proposed techniques might need adaptation. Potentially, off-screen visualization techniques could be integrated into an AR setting to compensate for this limited field of view \cite{gruenefeld2017visualizing}. Also, in AR displays without occlusion management, physical objects, such as the tablet or the user's hands are visible alongside the virtual objects, typically at a different focus distance, which can further lead to potential perceptual issues \cite{eiberger2019effects}. In contrast, combining an AR HMD with a physical tablet could make use of the potentially higher output resolution of the physical touch display and open up further possibilities for novel overview + detail techniques.



\subsubsection{Limitations}
Our work focused on a subset of a possibly large design space. \hl{While we foresee the combination of VR HMDs and tablets as one potentially promising direction on how mobile knowledge workers can be supported in the future, we are aware that it remains uncertain if and when such potential combinations will become products due to various factors such as technology, market, regulation, social acceptance and timing uncertainty \cite{jalonen2012uncertainty}}. For the \textsc{Content Transfer Task}, we tested both a small and a large number of virtual screens. However, other factors such as screen arrangement, screen size, or parameters of the interaction methods could impact the results. \hl{Although we foresee no substantial reversal of the effects indicated in the study, it may be interesting, as a future work, to test the limit of noisy eye gaze usage for very small screens and  translation distances, following} \cite{turner2015Gaze}.  For the \textsc{puzzle task} we empirically determined the design parameters, such as distance between layers. Future work should investigate the effects of layer distance on various task types in more detail, as with increasing layer distance, the association between individual $(x,y)$ coordinates on a layer could become harder for users to track. 
While we indicated the feasibility of mobile implementations using our prototype, we still opted for using an external tracking system for implementing our interaction techniques due to the limited accuracy of generation HMD-based hand tracking \cite{schneider2020accuracy} at the time of the study. Newer technologies, such as Oculus Quest hand tracking offer higher accuracy and may enable sufficient performance. It remains to be evaluated how robust concurrent camera-based tracking of the hands, keyboard, the touchscreen, and, potentially, a pen is in real-world usage scenarios. An alternative approach may use a tablet to record the user's in air hand movement using a 45-degree mirror clipped above the tablet front camera \cite{yoo2016Symm}.

\section{Conclusions and Future Work}


Our work explored the opportunities arising out of a joint interaction space between VR HMDs and laptop/tablet capacitive touchscreens for supporting mobile information workers. We have demonstrated that such a joint interaction space is feasible to be implemented using consumer-oriented hardware by tracking the touchscreen and the user's finger, which could be useful in small constrained environments such as on an airplane or in touchdown spaces. We explored how to spatially arrange and interact with data within this joint interaction space. We further implemented a set of six applications (window manager, code version control, parallel coordinates exploration, map navigation, medical image viewer, and a virtual PowerPoint) making use of this design space. In future work, these interaction concepts should be studied in depth, such as recently done for spreadsheet interaction \cite{Gesslein2020Pen}, to allow us to better understand their capabilities and limitations for supporting mobile knowledge workers in real-world working contexts. \hl{Also, additional study designs, e.g., for further investigations of the explosion diagram technique,  comparisons with non-VR baseline techniques, effects of physical and virtual screen sizes, as well as the use of fully mobile implementations in real-world environments seem feasible.} In particular, we see the potential for further multimodal interaction techniques that fit this shallow 3D space spanned by the hands resting on a 2D screen. One avenue of future work is to explore which particular document semantics are the most suitable for display above the documents (such as, for example, comments and other mark-up on a text document, notes on a code review), and which mappings fit different contexts. In addition, another avenue is to explore how to expand the interaction vocabulary with simultaneous pen and touch techniques. 




\balance{}


\bibliographystyle{abbrv-doi}

\bibliography{vrmobilework}

\end{document}